\documentclass{aa}

\newcommand{\degpt}[2]{\mbox{$\rm #1\hspace{-0.25em}\stackrel{\circ}{.}
      \hspace{-1.0mm}#2$}}                          
\newcommand{\magpt}[2]{\mbox{$\rm #1\hspace{-0.25em}\stackrel{m}{.}
      \hspace{-1.0mm}#2$}}                             
\newcommand{\magn}[1]{\mbox{$\rm #1\hspace{-0.05em}^m$}}
\def\bsec{\hbox{$.\!\!{\arcsec}$}}

\newcommand\RA[4]{#1$^{\rm h}$#2$^{\rm m}$#3$\stackrel{\rm s}{.}$#4}

\newcommand\DEC[4]{#1$^{\circ}$#2\arcmin#3\bsec#4}
\newcommand\teff{$ {\rm T_{eff}}$}

\newcommand\logg{$\log {\rm g}$}
\newcommand\loghe{${\rm \log{\frac{n_{He}}{n_{H}}}}$}

\newcommand\ebv{$ {\rm E_{B-V}}$}
\def\gtrsim{\mathrel{\hbox{\rlap{\hbox{\lower4pt\hbox{$\sim$}}}\hbox{$>$}}}}
\def\lesssim{\mathrel{\hbox{\rlap{\hbox{\lower4pt\hbox{$\sim$}}}\hbox{$<$}}}}

\newcommand{\Msolar}{\mbox{\,$\rm M_{\odot}$}}        
\begin{document}

\title{Spectroscopic Analyses of the Blue Hook Stars in NGC~2808:
A More Stringent Test of the Late Hot Flasher Scenario\thanks{Based on
observations collected at the European Southern Observatory, Chile
(ESO proposal 68.D-0248)} } 
\author{S. Moehler\inst{1} \and A.V.~Sweigart\inst{2} \and W.B.~Landsman\inst{3} \and N.J.~Hammer\inst{4} \and
S. Dreizler\inst{4,5}} 
\offprints{S. Moehler} 

\institute{
Institut f\"ur Theoretische Physik und Astrophysik der Universit\"at
Kiel, Abteilung Astrophysik, 24098 Kiel, Germany (e-mail:
moehler@astrophysik.uni-kiel.de)
\and NASA\,Goddard Space Flight Center, Code 681, Greenbelt, MD 20771,
USA (e-mail: Allen.V.Sweigart@nasa.gov)
\and SSAI, NASA\,Goddard Space Flight Center, Code 681, Greenbelt, MD
20771, USA (e-mail: landsman@mpb.gsfc.nasa.gov)
\and Institut f\"ur Astronomie und Astrophysik
der Universit\"at T\"ubingen, Sand 1,
D-72076 T\"ubingen, Germany (e-mail: hammer@astro.uni-tuebingen.de)
\and Universit\"ats-Sternwarte G\"ottingen, Geismarlandstr. 11,
D-37083 G\"ottingen (e-mail: dreizler@astro.physik.uni-goettingen.de)}
\titlerunning{Blue Hook Stars in NGC~2808} 
\authorrunning{Moehler, Sweigart, Landsman, Hammer, Dreizler}
\date{Received 14 October 2003, accepted 4 November 2003} 

\abstract{Recent UV observations of the globular cluster NGC~2808
(Brown et al. \cite{brsw01}) show a significant population of hot
stars fainter than the zero-age horizontal branch (``blue hook''
stars), which cannot be explained by canonical stellar evolution.
Their results suggest that stars which experience unusually large mass
loss on the red giant branch and which subsequently undergo the helium
core flash while descending the white dwarf cooling curve could
populate this region.  Theory predicts that these ``late hot
flashers'' should show higher temperatures than the hottest canonical
horizontal branch stars and should have helium- and carbon-rich
atmospheres. As a test of this late hot flasher scenario, we have
obtained and analysed medium resolution spectra of a sample of blue
hook stars in NGC~2808 to derive their atmospheric parameters.  Using
the same procedures, we have also re-analyzed our earlier spectra of
the blue hook stars in $\omega$ Cen (Moehler et al. \cite{mosw02}) for
comparison with the present results for NGC~2808.  The blue hook stars
in these two clusters are both hotter (\teff $\ge$35,000~K) and more
helium-rich than canonical extreme horizontal branch stars in
agreement with the late hot flasher scenario. Moreover, we find
indications for carbon enhancement in the three most helium-enriched
stars in NGC~2808.  However, the blue hook stars still show some
hydrogen in their atmospheres, perhaps indicating that some residual
hydrogen survives a late hot flash and then later diffuses to the
surface during the horizontal branch phase. We note that the presence
of blue hook stars apparently depends mostly on the total mass of the
globular cluster and not so much on its horizontal branch morphology.
} \maketitle

\section{Introduction}
\begin{figure}[h]
\vspace*{9.cm}
\includegraphics{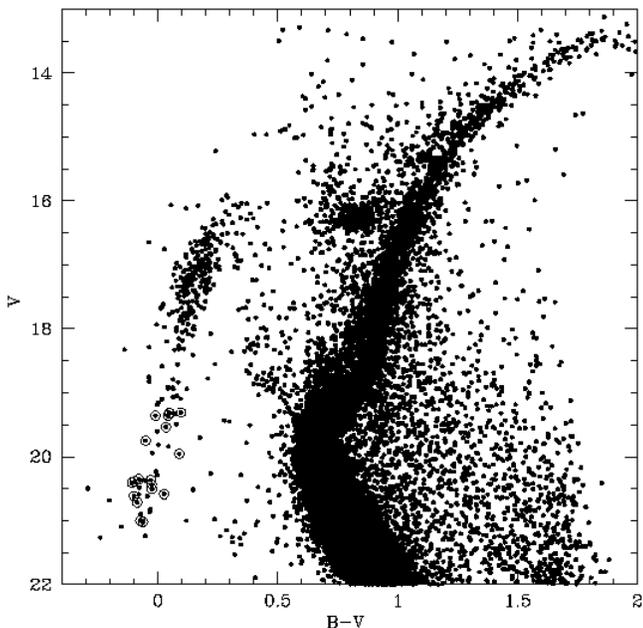}
\caption{
Colour-magnitude diagram of NGC~2808 (Walker \cite{walk99}) with
our spectroscopic targets marked.  Our targets were chosen to span
the gap at V = \magpt{20}{0} which, according to the hot flasher
scenario, separates the canonical EHB stars from the fainter blue hook
stars.
\label{cmd}}

\end{figure}

Low-mass stars burning helium in a core of about 0.5~\Msolar\ and
hydrogen in a shell populate a roughly horizontal region in the
colour-magnitude diagrams of globular clusters, which has earned them
the name ``horizontal branch" (HB) stars.  The Galactic globular
clusters show a great variety in horizontal branch morphology, i.e., in
the temperature distribution of their HB stars.  The temperature of an
HB star depends -- at a given metallicity -- on the mass of its hydrogen
envelope, with the hottest or extreme HB (EHB) stars (\teff\ $>$
20,000~K) having extremely thin  $\lesssim$0.01\Msolar) envelopes. The
increase in the bolometric correction with increasing temperature
turns the blue HB into a vertical blue tail in optical
colour-magnitude diagrams with the faintest blue tail stars being the
hottest and least massive (see Fig.~\ref{cmd} for a prominent example
of a blue tail). It is still unclear how some stars manage to lose
nearly all of their envelope mass and still undergo the helium core flash,
as indicated by the long blue tails in a number of globular clusters (e.g.,
NGC~6752, Moehler et al. \cite{mosw00}). The globular cluster
NGC~2808 represents a good template cluster, 
since it has an extremely long blue tail extending to very
high temperatures and correspondingly low envelope masses.

NGC 2808 has been the subject of four recent deep ground-based and Hubble Space Telescope (HST) photometric studies (Sosin et
al.\ \cite{sosi97}, Walker \cite{walk99}, Bedin et al.\
\cite{bepi00}, Brown et al.\ \cite{brsw01}).  The ground-based
studies have shown that the blue tail of NGC~2808 extends to extremely
faint magnitudes ($M_V \approx$ \magpt{5}{5}), with gaps at $M_V
\approx$ \magn{3} and \magpt{4}{5}. These gaps can be seen
at V $\approx \magpt{18}{5}$ and $\magpt{20}{0}$ in
Fig.~\ref{cmd}. The brighter gap is thus at a similar position as
the underpopulated region in the colour-magnitude diagram of NGC~6752,
which separates the EHB stars from the classical blue HB
stars (Moehler et al.  \cite{mohe97b}, \cite{mosw00}; Momany et
al. \cite{mopi02}), while the fainter gap coincides with the hot end
of the blue tail in NGC~6752.  The blue tail in NGC~2808 thus extends
to hotter temperatures than that of NGC~6752.  That in itself would
not be a problem, but the spectroscopic analyses of Moehler et al.
(\cite{mohe97b}, \cite{mosw00}) show that the blue tail stars in
NGC~6752 already populate the EHB to the hot end predicted by
canonical HB models.

Observations of NGC~2808 in the far- and near-UV (Brown et
al. \cite{brsw01}, their Fig.\ 3) show that the stars below the
faint gap form a hook-like feature that extends up to
\magpt{0}{7} below the hot end of the zero-age HB (ZAHB). Such ``blue
hook'' stars are so far known in only three other globular
clusters: $\omega$ Cen (D'Cruz et al. \cite{dcoc00}, Moehler et
al. \cite{mosw02}), NGC~6388 and possibly NGC~6441 (Busso et
al. \cite{bupi03}), all of which are very massive.  Within the
framework of canonical HB theory there is no way to populate this
region of the UV colour-magnitude diagram without requiring an
implausibly large decrease in the helium-core mass.  Thus canonical
theory fails to explain both the faint UV luminosities and expected
high temperatures of the blue hook stars.

Brown et al.\ (\cite{brsw01}) have proposed a scenario to explain the
blue hook stars.  According to this scenario the stars brighter than
the gap at $M_V \approx$ \magpt{4}{5} are canonical EHB stars
which lost all but $\sim$0.01\Msolar\ of their envelope mass prior to
undergoing the helium core flash.  As discovered by Castellani \&
Castellani (\cite{caca93}), stars which lose more mass than this will
leave the red giant branch and evolve to high effective temperatures
before undergoing the helium core flash, producing the so-called ``hot
flashers".  Sweigart (\cite{swei97}) subsequently showed that hot
helium flashes can occur either as a star evolves from the tip of the
red giant branch to the top of the white dwarf cooling curve (``early"
hot flasher), or later as a star descends the white dwarf cooling
curve (``late" hot flasher).  For even higher mass loss, a star will
die as a helium white dwarf without ever igniting helium.  These
different evolutionary paths are illustrated in Moehler et
al. (\cite{mosw02}) and Sweigart et al. (\cite{swbr02}).

D'Cruz et al. (\cite{dcdo96}, \cite{dcoc00}) proposed that the blue
hook stars could be the progeny of such hot flashers, but
unfortunately the D'Cruz et al. models were, at most, only
$\approx$\magpt{0}{1} fainter than the canonical ZAHB, much less than
required by the observations.  More recently, Brown et
al. (\cite{brsw01}) have explored the evolution of both the early and
late hot flashers through the helium core flash to the EHB in more
detail.  Their models show that a late hot helium flash on the white
dwarf cooling curve will induce substantial mixing between the
hydrogen envelope and helium core, leading to helium-rich EHB stars
that are much hotter than canonical ones, as found previously by
Sweigart (\cite{swei97}).  This result has been confirmed by the
calculations of Cassisi et al.  (\cite{casc03}), who were able to
follow the evolution of a late hot flasher completely through the
helium core flash.  Brown et al.\ (\cite{brsw01}) suggest that such
flash mixing may be the key for understanding the evolutionary status
of the blue hook stars. Such mixing may also be responsible for
producing the helium-rich, high gravity field sdO stars (Lemke et
al. \cite{lehe97}), whose origin is otherwise obscure.  The flash
mixing scenario predicts a gap of about 6000~K between the
canonical EHB stars (i.e., stars without flash-mixing, including
early hot flashers) and the late hot flashers as well as a helium
dominated atmospheric composition for the late hot flashers. The
hydrogen deficiency results in a flux redistribution where more flux
is emitted shortward of the Lyman edge at 912 \AA\, and less flux is
emitted at ultraviolet (1500 \AA) and longer wavelengths, which makes
flash-mixed stars fainter than canonical EHB stars at these
wavelengths.

Spectroscopic observations of blue hook stars in $\omega$ Cen by Moehler et al.
(\cite{mosw02}) showed that these stars reach effective temperatures of more
than 35,000~K, i.e.,  well beyond the hot end of the canonical EHB.  In
addition, most of them show at least solar helium abundances with the helium
abundance increasing with effective temperature, in contrast to canonical EHB
stars such as those studied in NGC~6752 by Moehler et al. (\cite{mosw00}). 
Contrary to the predictions of Brown et al. (\cite{brsw01}), however, only one
star showed a helium abundance of \loghe\ $>$ 0. These results may indicate
that flash mixing is less efficient than assumed, or that some
residual hydrogen survives flash mixing and then diffuses outward to the
surface. It is unclear whether the lower than expected helium abundances can
still reproduce the UV properties of the stars.  UV data have been
presented for the blue hook stars in the central region of $\omega$ Cen by
D'Cruz et al. (\cite{dcoc00}),  but due to crowding these stars are not
accessible to ground-based spectroscopy.   UV data are available for the blue
hook stars in the outer region of $\omega$ Cen from the Ultraviolet Imaging
Telescope (UIT, Stecher et al. \cite{stec97}), but  unfortunately, these data
are too noisy to permit a quantitative test. 

Considering these
somewhat ambiguous results and the fact that $\omega$ Cen is not a
typical globular cluster, we decided to test the late hot flasher
scenario by observing  the more typical globular cluster NGC~2808.
We discuss our observational data in Sect.~\ref{sec-observ} and then
derive the parameters of the blue hook stars (temperatures, gravities
and helium abundances) in Sect.~\ref{sec-analysis}.  In
Sect.~\ref{sec-discuss} we compare our results with the predictions of
the flash-mixing scenario.

\section{Observations and Data Reduction\label{sec-observ}}

\subsection{Target Selection}
 Our spectroscopic targets were selected from the catalog of
Walker (\cite{walk99}; Fig.\ref{cmd}), and include seven stars
brighter than the gap at V $\approx$ \magn{20}, and twelve stars
fainter than the gap.  The coordinates and photometry for our targets
are given in Table~\ref{ngc2808_par}, along with cross-identifications
from the catalog of Bedin et al. (\cite{bepi00}).

\begin{table*}
\caption{Positions and photometric information of target
stars\label{ngc2808_par}}
\begin{tabular}{ll|llllll|l}
\hline
\hline
 $\alpha_{2000}$ & $\delta_{2000}$ & \multicolumn{3}{c}{Walker (1999)}
 & \multicolumn{3}{c|}{Bedin et al. (2000)} & m$_{1520}$\\
 & & & $V$ & $B-V$ & & $V$ & $B-V$ & \\
\hline
\multicolumn{9}{c}{\bf setup a}\\
\hline
\RA{09}{12}{07}{20} & \DEC{$-$64}{54}{41}{1} & W6816 & \magpt{19}{35} & 
 \magpt{$+$0}{04} & B13283 & \magpt{19}{38} & \magpt{-0}{00} &
 \magpt{16}{51}\\
\RA{09}{12}{20}{68} & \DEC{$-$64}{53}{09}{4} & W9301 & \magpt{19}{32} & 
 \magpt{$+$0}{05} & B12148 & \magpt{19}{27} & \magpt{$+$0}{08} &
 \magpt{16}{63}\\
\RA{09}{11}{57}{06} & \DEC{$-$64}{56}{18}{6} & W14040 & \magpt{19}{96} & 
 \magpt{$+$0}{09}  & B12955 & \magpt{19}{95} &  \magpt{$+$0}{03} &
 \magpt{16}{75}\\
\RA{09}{11}{58}{81} & \DEC{$-$64}{56}{45}{6} & W14711 & \magpt{19}{31} & 
 \magpt{$+$0}{10}  & B12583 & \magpt{19}{36} & \magpt{-0}{02} & 
 \magpt{16}{48}\\
\RA{09}{12}{09}{22} & \DEC{$-$64}{54}{08}{3} & W18803 & \magpt{20}{72} & 
 \magpt{-0}{09} & B20678 & \magpt{20}{60} &  \magpt{$+$0}{14} & 
 \magpt{16}{96}\\
\RA{09}{12}{09}{47} & \DEC{$-$64}{55}{03}{5} & W18899 & \magpt{21}{00} & 
 \magpt{-0}{07} & B34817 & \magpt{21}{06} & \magpt{-0}{08} & 
 \magpt{17}{01}\\
\RA{09}{12}{20}{95} & \DEC{$-$64}{53}{57}{2} & W23198 & \magpt{20}{50} & 
 \magpt{-0}{03} & B21073 & \magpt{20}{48} & \magpt{-0}{08} & 
 \magpt{16}{71}\\
\RA{09}{12}{28}{51} & \DEC{$-$64}{52}{19}{3} & W25794 & \magpt{20}{41} & 
 \magpt{-0}{11} & B21376 & \magpt{20}{32} &  \magpt{$+$0}{01} & 
 \magpt{16}{71}\\
\RA{09}{12}{31}{25} & \DEC{$-$64}{52}{06}{9} & W26607 & \magpt{20}{58} & 
 \magpt{$+$0}{03}  & B31854 & \magpt{20}{71} &  \magpt{$+$0}{00} & 
 --\\
\hline
\multicolumn{9}{c}{\bf setup b}\\
\hline
\RA{09}{12}{13}{42} & \DEC{$-$64}{46}{35}{4} & W20334 & \magpt{21}{01} & 
 \magpt{-0}{06} & B38604 & \magpt{21}{02} & \magpt{-0}{05} & 
 \magpt{17}{35}\\
\RA{09}{12}{17}{43} & \DEC{$-$64}{48}{03}{5} & W21882 & \magpt{19}{75} & 
 \magpt{-0}{05} & B14334 & \magpt{19}{70} & \magpt{-0}{04} & 
 \magpt{15}{38}\\
\RA{09}{12}{29}{81} & \DEC{$-$64}{47}{56}{1} & W26182 & \magpt{19}{32} & 
 \magpt{$+$0}{05}  & B11516 & \magpt{19}{33} &  \magpt{$+$0}{03} &
 \magpt{16}{56}\\
\RA{09}{12}{34}{11} & \DEC{$-$64}{51}{23}{2} & W27412 & \magpt{19}{54} & 
 \magpt{$+$0}{03}  & B15455 & \magpt{19}{64} & \magpt{-0}{00} &
 \magpt{16}{32} \\
\hline
\multicolumn{9}{c}{\bf setup c}\\
\hline
\RA{09}{11}{33}{72} & \DEC{$-$64}{56}{24}{9} & W6022 & \magpt{19}{36} & 
 \magpt{-0}{01} & B11345&  \magpt{19}{50} & \magpt{-0}{12} &
 \magpt{15}{93}\\
\RA{09}{11}{36}{59} & \DEC{$-$64}{58}{01}{0} & W6849 & \magpt{20}{35} & 
 \magpt{-0}{08} & B23335&  \magpt{20}{35} & \magpt{-0}{13} & 
 \magpt{16}{98}\\
\RA{09}{11}{37}{53} & \DEC{$-$64}{53}{31}{1} & W7084 & \magpt{20}{44} & 
 \magpt{-0}{03} & B26161 & \magpt{20}{59} & \magpt{-0}{03} &
 \magpt{16}{82}\\
\RA{09}{11}{42}{47} & \DEC{$-$64}{56}{57}{1} & W8750 & \magpt{20}{62} & 
 \magpt{-0}{10} & B25082&  \magpt{20}{60} & \magpt{-0}{14} &
 \magpt{16}{87}\\
\RA{09}{11}{36}{34} & \DEC{$-$64}{54}{07}{7} & W9855 & \magpt{20}{36} & 
 \magpt{-0}{03} & B24578 & \magpt{20}{50} & \magpt{-0}{15} & 
 \magpt{16}{66}\\
\RA{09}{11}{45}{79} & \DEC{$-$64}{56}{27}{1} & W9863 & \magpt{20}{39} & 
 \magpt{-0}{10} & B16326&  \magpt{20}{36} & \magpt{-0}{14} & 
 \magpt{16}{20} \\
\hline
\end{tabular}
\end{table*}

\subsection{UIT observations}

\begin{figure}[h]
\vspace*{9.cm}
\includegraphics{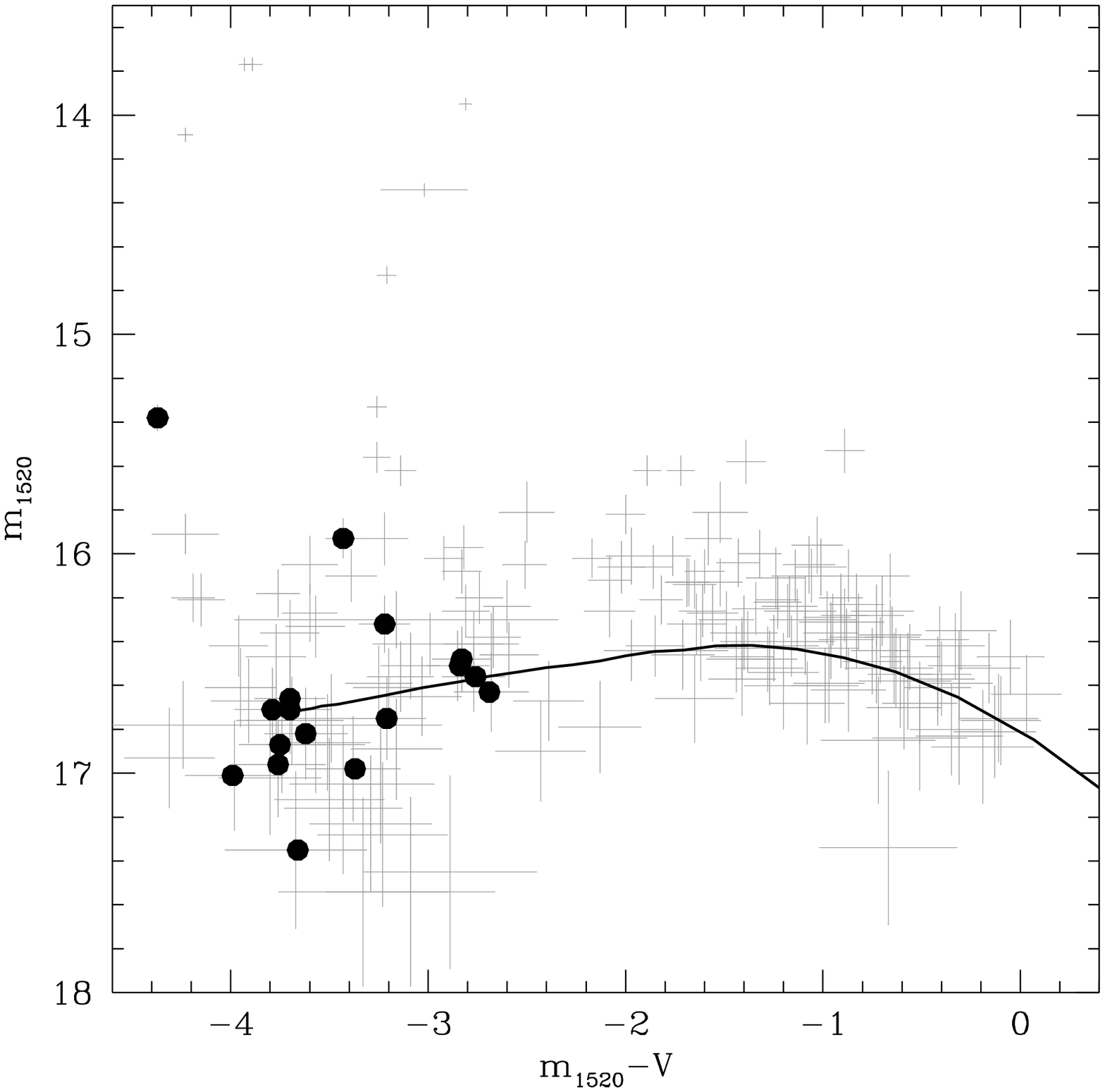}
\caption{Ultraviolet-visual colour-magnitude diagram of NGC~2808 as
derived from UIT observations. The spectroscopic targets are marked by
big dots. The zero-age HB (solid line) is computed for [M/H] = $-$1.1
and has been transformed assuming a distance modulus (m-M)$_0$ of
\magpt{15}{04} and a reddening of \ebv\ = \magpt{0}{18} as
suggested by Bedin et al. (\cite{bepi00}) for the Zinn \& West
metallicity scale. If we adopt the distance and reddening determined
by Bedin et al. (\cite{bepi00}) for the Caretta \& Gratton scale, then
the ZAHB would be $\approx$\magpt{0}{2} fainter.\label{uitcmd}}
\end{figure}

We would like to compare the results of our ground-based
spectroscopy with the positions of our target stars in the UV
colour-magnitude diagram.  Unfortunately, we cannot use the UV
colour-magnitude diagram of Brown et al. (\cite{brsw01}) because all
of their stars lie within 35\arcsec\ of the cluster center and are
thus too crowded for ground-based spectroscopy.  As an alternative, we
have derived new 1520~\AA\ photometry of stars in the outer region of
NGC~2808 using archival UIT images.  The UIT photometry of NGC~2808
has not been previously published, although optical spectroscopy of
three UV-bright stars on the image was presented by Moehler et
al. (\cite{mola98}).  The UIT image of NGC~2808 was a 979 second
exposure obtained on 19 May 1995 using the B1 filter, which has
a central wavelength of 1520~\AA\ and a width of 350~\AA.  The FWHM of
the star images was about 5\arcsec\ and the faintest hot HB stars were
near the sensitivity limit.  Therefore, to create an ultraviolet
colour-magnitude diagram, we performed 3-pixel (3\farcs4) radius
circular aperture photometry only on isolated stars (all more than
2\arcmin\ from the cluster center) which had a positional match with
hot ($B-V <$ \magpt{0}{2}) stars in the catalogs of Walker
(\cite{walk99}) or Bedin et al. (\cite{bepi00}). A star was determined
to be isolated if it had no neighbours within 5\arcsec\ on either the
UIT image, or on the lists of hot stars in the optical catalogs.  We
used the standard UIT absolute calibration and determined an aperture
correction of \magpt{0}{94} using a few isolated UV-bright stars. The resulting UIT colour-magnitude diagram obtained from the
1520~\AA\ photometry given in Table~\ref{ngc2808_par} is shown in
Fig.~\ref{uitcmd}, and includes all our spectroscopic targets except
for W26607 which has a blurred PSF on the UIT image.  The complete
ultraviolet photometry and optical identifications used in
Fig.~\ref{uitcmd} are available upon request to the authors.

Although the UIT colour-magnitude diagram is of poorer quality than
that shown in Brown et al. (\cite{brsw01}), it is still possible to
identify the candidate sub-ZAHB stars.  Unfortunately, all but
one of our blue hook targets are less than 0.3 mag fainter than the
hottest canonical ZAHB model at m$_{1520}$ = 16.72.  Only W20334 at
m$_{1520}$ = 17.35 is near the faint end of the blue hook distribution
found by Brown et al. (2001).  Thus our choice of targets is biased
toward the brighter blue hook stars.

\subsection{Spectroscopy}
 We obtained medium-resolution spectra ($R\approx$900) of 19 stars
along the blue tail in NGC~2808 with \magn{19} $< V <$ \magn{21} at
the VLT-UT1 (Antu) with FORS1  see Fig.~\ref{cmd}). 
The data were obtained in service mode
(see Table~\ref{tab-obs} for details).  We used the multi-object
spectroscopy (MOS) mode of FORS1 (slit length 19\arcsec) with the
standard collimator (0\bsec2/pixel), a slit width of 1\arcsec\ and
grism B600. As FORS1 is equipped with an atmospheric dispersion
corrector, MOS observations at higher airmass are not a problem. The
detector was a TK2048EB4-1 backside thinned CCD with 2048$\times$2048
pixels of (24 $\mu$m)$^2$, which was read out with high gain (1.46
e$^-$/count, 5.15 e$^-$ read-out noise) and normal read-out speed
using one read-out port only. This configuration yields a dispersion
of 1.2~\AA/pixel. As can be seen from Table~\ref{tab-obs}, the seeing
was usually better than the slit width, resulting in a
seeing-dependent resolution of the spectra.

\begin{table*}
\caption[]{Observational parameters\label{tab-obs}}
\begin{tabular}{lrrrrr}
\hline
\hline
setup & start of observation & seeing & airmass & \multicolumn{2}{c}{moon}\\
&  &  &  & illumination & 
distance\\
\hline
a & 11/01/2002 05:38:52.722 & 0\bsec91 & 1.319 & 0.053 & \degpt{83}{3} \\  
  & 11/01/2002 06:37:32.265 & 0\bsec61 & 1.319 & 0.050 & \degpt{83}{4} \\  
  & 15/01/2002 05:13:58.529 & 0\bsec73 & 1.322 & 0.027 & \degpt{95}{3} \\  
b & 20/01/2002 04:50:50.581 & 0\bsec77 & 1.322 & 0.350 & \degpt{105}{0} \\
  & 08/02/2002 02:26:30.563 & 0\bsec82 & 1.390 & 0.161 & \degpt{85}{0} \\  
  & 10/02/2002 02:46:38.787 & 0\bsec68 & 1.354 & 0.046 & \degpt{91}{2} \\  
c & 09/02/2002 04:02:45.640 & 0\bsec67 & 1.314 & 0.090 & \degpt{88}{2} \\  
  & 09/02/2002 04:46:30.947 & 0\bsec50 & 1.318 & 0.088 & \degpt{88}{3} \\
\hline
\end{tabular}
\end{table*}

For each night dome flat fields with two different illumination
patterns and CdHeHg wavelength calibration spectra were observed. As
part of the standard calibration we were also provided with masterbias
frames for our data. The masterbias showed no evidence for hot pixels
and was smoothed with a 30$\times$30 box filter to keep any possible
large scale variations, but erase noise. The flat fields were averaged
for each night and bias-corrected by subtracting the smoothed
masterbias of that night. From the flat fields we determined the
limits of the slitlets in spatial direction. Each slitlet was
extracted and from there on treated like a long-slit spectrum. The
flat fields were normalized with 4$^{\rm th}$ or 5$^{\rm th}$ order
polynomials. The dispersion relation was obtained from the wavelength
calibration frames by fitting 3$^{\rm rd}$ to 4$^{\rm th}$ order
polynomials to the line positions along the dispersion axis. We used
13 to 16 unblended lines between 3600~\AA\ and 6200~\AA\ and achieved
an r.m.s. error of 0.05~\AA\ to 0.08~\AA\ per row.

Due to exposure times of 2700 sec the scientific observations
contained a large number of cosmic ray hits. Those were corrected with
the algorithm described in G\"ossl \& Riffeser (\cite{gori02}), using
a threshold of 15$\sigma$ and a FWHM of 1.5 pixels for the cosmic ray
hits. As the routine is not originally intended for the use with
spectra, we also reduced the uncorrected frames to allow a check for
any possible artifacts of the cosmic ray cleaning procedure.  The
slitlets with the stellar spectra were extracted in the same way as
the flat field and wavelength calibration slitlets. The smoothed
masterbias was subtracted, and the spectra were divided by the
corresponding normalized flat fields, before they were rebinned
2-dimensionally to constant wavelength steps.  We then filtered the
uncleaned frames along the spatial axis with a median filter of 7
pixels width to erase cosmic ray hits. We identified regions
uncontaminated by any stellar source and approximated the spatial
distribution of the sky background by a constant. For the uncleaned
frames the sky background was fit on the median filtered frames and
subtracted from the unfiltered ones, whereas for the cleaned frames
both sky fit and subtraction were performed on the unfiltered rebinned
data.  The sky-subtracted spectra were extracted using Horne's
(\cite{horn86}) algorithm as implemented in MIDAS.  Finally the
spectra were corrected for atmospheric extinction using the extinction
coefficients for La Silla (T\"ug \cite{tueg77}) as implemented in
MIDAS, because they provide the closest approximation to Paranal
conditions, for which no spectroscopic extinction coefficients are
available.

For a relative flux calibration we used response curves derived from
spectra of LTT~3864, LTT~4364, and LTT~6248 with the data of Hamuy et
al. (\cite{hawa92}). The response curves were fit by splines and
averaged for all nights.  The individual target spectra were
corrected for any Doppler shifts determined from Balmer and/or helium
absorption lines.  These shifts varied more than expected from one
observation to the next. This can be understood by the fact that the
seeing disk was smaller than the slit width, so that small shifts in
the position of the stars within the slit can introduce shifts in the
wavelength. A shift of 0\bsec1 corresponds to a shift of 0.6~\AA.  Any
remaining cosmic ray hits in the stellar spectra were corrected by
comparing the individual spectra of the stars. Due to the seeing being
smaller than the slit width spectra of the same star from different
nights can have different resolution.  In order to get a well defined
resolution of the co-added spectra, we convolved spectra obtained with
better seeing to the resolution determined by the worst resolution of
each setup. This resulted in resolutions of 5.4~\AA\ (0\bsec9) for
setup a, 4.8~\AA\ (0\bsec8) for setup b and 4.0~\AA\ (0\bsec7) for
setup c.

\begin{figure*}[ht]
\vspace*{20cm}
\includegraphics{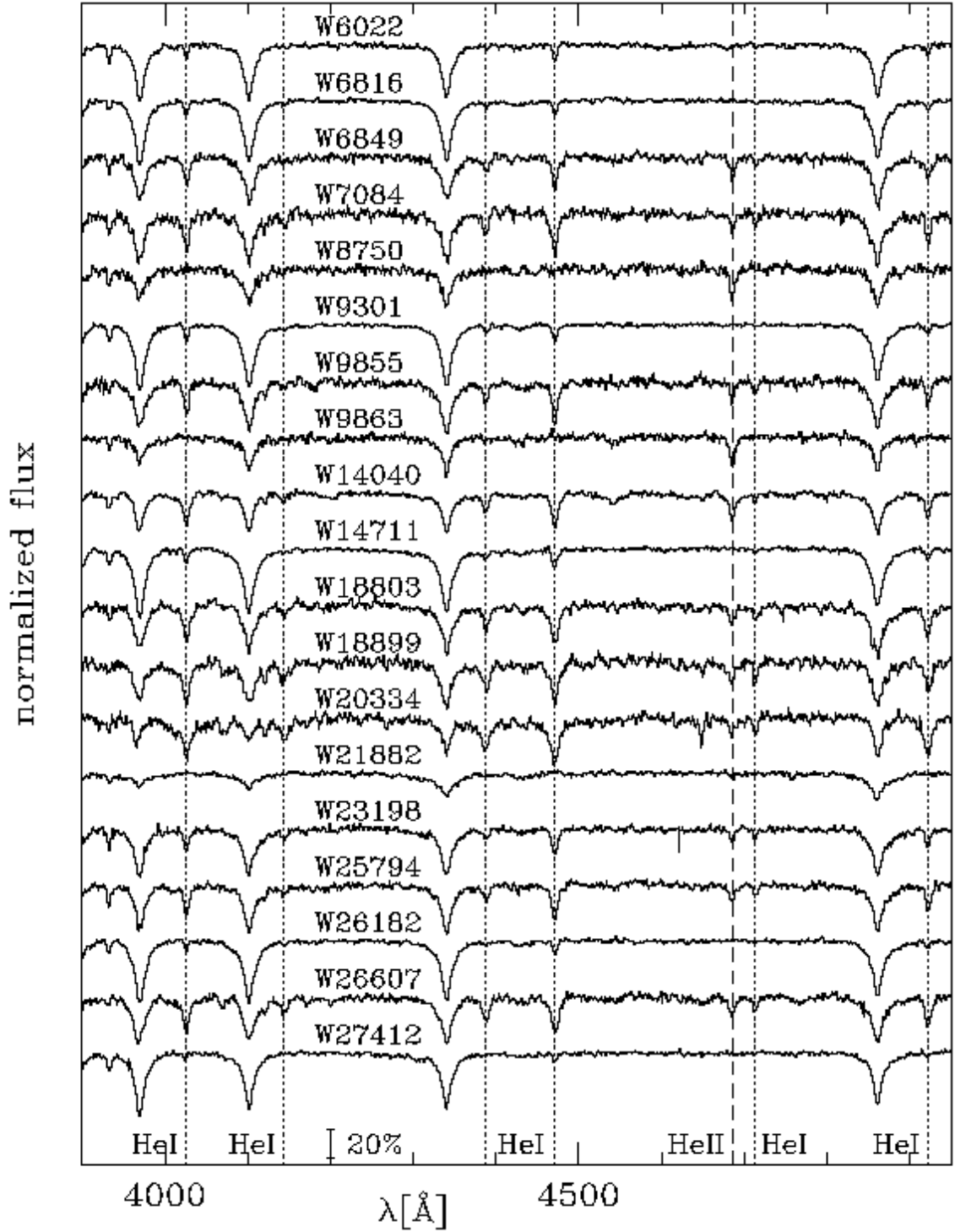}
\caption{Spectra of blue tail stars in NGC~2808 sorted by their number
in the photometry of Walker (\cite{walk99}). The spectra were
normalized by eye to allow a better comparison. The lines mark the
position of the He~{\sc i} and He~{\sc ii} lines.
\label{spec}}
\end{figure*}

\section{Analysis\label{sec-analysis}}

The spectra, given in Fig.~\ref{spec}, show a large variety of
helium line strengths. Part of this variation is due to variations in
effective temperature, but -- as we will see in the analysis --
part is also due to variations in the atmospheric helium abundance
along the blue tail in NGC~2808.

\subsection{Atmospheric parameters}

\begin{table}
\caption[]{Atmospheric parameters of the stars in NGC~2808\label{tab-par}}
\begin{tabular}{llrrr}
\hline
\hline
Star & $\chi^2$ & \teff & \logg & \loghe\\
     &          & [K]   &       &       \\
\hline
W6022 & 2.23 & 30300$\pm$\hspace*{0.5ex} 480 & 5.57$\pm$0.05 & 
 $-$1.99$\pm$0.07 \\
W6816 & 2.39 & 21300$\pm$\hspace*{0.5ex} 540 & 4.88$\pm$0.07 & 
 $-$2.17$\pm$0.05 \\
W6849 & 1.88 & 37500$\pm$\hspace*{0.5ex} 330 & 5.77$\pm$0.09 & 
 $-$0.82$\pm$0.05 \\
W7084 & 1.63 & 35300$\pm$\hspace*{0.5ex} 580 & 5.91$\pm$0.10 & 
 $-$0.57$\pm$0.05 \\
W8750 & 2.52 & 48000$\pm$1300 & 6.21$\pm$0.14 & 
 $-$1.78$\pm$0.15 \\
W9301 & 2.14 & 21100$\pm$\hspace*{0.5ex} 560 & 4.92$\pm$0.07 & 
 $-$2.05$\pm$0.05 \\
W9855 & 2.45 & 35700$\pm$\hspace*{0.5ex} 580 & 5.60$\pm$0.12 & 
 $-$0.74$\pm$0.05 \\
W9863 & 2.04 & 53000$\pm$2100 & 5.81$\pm$0.10 & 
 $-$1.56$\pm$0.14 \\
W14040 & 2.16 & 40400$\pm$\hspace*{0.5ex} 220 & 5.63$\pm$0.05 & 
 $-$0.31$\pm$0.03 \\
W14711 & 2.92 & 20100$\pm$\hspace*{0.5ex} 520 & 4.87$\pm$0.07 & 
 $-$1.93$\pm$0.05 \\
W18803 & 1.70 & 35700$\pm$\hspace*{0.5ex} 510 & 5.96$\pm$0.09 & 
 $-$0.38$\pm$0.03 \\
W18899 & 1.91 & 35500$\pm$\hspace*{0.5ex} 790 & 5.70$\pm$0.12 & 
 $-$0.21$\pm$0.05 \\
W20334 & 2.15 & 36300$\pm$\hspace*{0.5ex} 900 & 5.79$\pm$0.15 & 
 $+$1.02$\pm$0.02 \\
W21882 & 4.09 & 67000$\pm$1500 & 6.99$\pm$0.09 & 
 $-$2.91$\pm$0.12 \\
W23198 & 2.38 & 34600$\pm$\hspace*{0.5ex} 400 & 5.81$\pm$0.07 & 
 $-$1.22$\pm$0.05 \\
W25794 & 1.44 & 36600$\pm$\hspace*{0.5ex} 460 & 5.67$\pm$0.09 & 
 $-$0.65$\pm$0.03 \\
W26182 & 1.79 & 24200$\pm$\hspace*{0.5ex} 640 & 5.22$\pm$0.07 & 
 $-$2.37$\pm$0.05 \\
W26607 & 2.91 & 36300$\pm$\hspace*{0.5ex} 520 & 5.63$\pm$0.09 & 
 $-$0.36$\pm$0.03 \\
W27412 & 2.64 & 28700$\pm$\hspace*{0.5ex} 640 & 5.37$\pm$0.07 & 
 $-$2.77$\pm$0.12 \\
\hline
\end{tabular}
\end{table}

\begin{table}
\caption[]{Atmospheric parameters of the stars in $\omega$ Cen 
\label{tab-paro}}
\begin{tabular}{llrrr}
\hline
\hline
Star & $\chi^2$ & \teff & \logg & \loghe \\
     &          & [K]   &       &        \\
\hline
WF3-1 & 1.34 & 36000$\pm$2100 & 5.97$\pm$0.40 & 
 $-$2.07$\pm$0.50\\
BC6022 & 1.96 & 46000$\pm$1100 & 6.14$\pm$0.15 & 
 $-$1.76$\pm$0.14\\
BC8117 & 1.65 & 30000$\pm$1600 & 5.43$\pm$0.21 & 
 $-$2.33$\pm$0.26\\
BC21840 & 2.94 & 36000$\pm$1000 & 5.44$\pm$0.19 & 
 $-$0.78$\pm$0.12\\
C521 & 2.20 & 35000$\pm$\hspace*{0.5ex} 620 & 5.86$\pm$0.12 & 
$-$0.88$\pm$0.07\\
D4985 & 2.50 & 38000$\pm$\hspace*{0.5ex} 700 & 5.79$\pm$0.17 & 
 $-$0.80$\pm$0.14\\
D10123 & 2.40 & 35200$\pm$\hspace*{0.5ex} 690 & 5.73$\pm$0.14 & 
 $-$0.84$\pm$0.09\\
D10763 & 2.85 & 36000$\pm$1700 & 4.82$\pm$0.15 & 
 $+$0.50$\pm$0.03\\
D12564 & 1.75 & 36800$\pm$\hspace*{0.5ex} 960 & 5.73$\pm$0.17 & 
 $-$0.50$\pm$0.05\\
D14695 & 2.16 & 38800$\pm$\hspace*{0.5ex} 230 & 5.31$\pm$0.12 & 
 $-$0.16$\pm$0.05\\
D15116 & 1.50 & 40000$\pm$1700 & 6.08$\pm$0.38 & 
 $-$0.51$\pm$0.17\\
D16003 & 1.45 & 36600$\pm$\hspace*{0.5ex} 550 & 5.89$\pm$0.12 & 
 $-$1.00$\pm$0.09\\
\hline
\end{tabular}
\end{table}

We fitted the hot stars (\teff $\gtrsim$33,000~K) with H-He
non-LTE model atmospheres to derive effective temperatures, surface
gravities, and helium abundances.  The helium-rich non-LTE model
atmospheres were calculated with a modified version of the accelerated
lambda iteration code of Werner \& Dreizler (\cite{wedr99}). The model
atoms for hydrogen and helium as well as the handling of the line
broadening for the spectrum synthesis are similar to those of Werner
(\cite{wern96}).  The calculation of the helium-poor non-LTE model
atmospheres is described in Napiwotzki (\cite{napi97}). For the cooler
stars we used ATLAS9 model atmospheres for solar metallicity (Kurucz
\cite{kuru93}) to account for effects of radiative levitation (see
Moehler et al. \cite{mosw00} for details), from which we calculated
spectra with Lemke's version\footnote{For a description see
http://a400.sternwarte.uni-erlangen.de/$\sim$ai26/linfit/linfor.html}
of the LINFOR program (developed originally by Holweger, Steffen, and
Steenbock at Kiel University). To establish the best fit, we used the
routines developed by Bergeron et al.\ (\cite{besa92}) and Saffer et
al.\ (\cite{sabe94}), as modified by Napiwotzki et
al. (\cite{nagr99}), which employ a $\chi^2$ test. The $\sigma$
necessary for the calculation of $\chi^2$ is estimated from the noise
in the continuum regions of the spectra. The fit program normalizes
model spectra {\em and} observed spectra using the same points for the
continuum definition. During the analysis of the data presented here
we realized that helium-poor and helium-rich spectra required
different sets of continuum points. Especially the use of the
continuum points derived from helium-poor spectra for the analysis of
helium-rich spectra can introduce large systematic errors,
esp. overestimates of the helium abundance due to continuum points in
the wings of strong helium lines and/or too narrow fitting windows for
helium lines. We therefore refined the definition of the continuum
points to correct for such errors.  Using this new
definition of the continuum points, we re-analysed the data for the
blue hook stars in $\omega$ Centauri presented in Moehler et
al. (\cite{mosw02}).  The new results differ from the old ones usually
within the mutual error bars, except for the two most helium-rich
stars, as expected.

We used the Balmer lines H$_\beta$ to H$_{8}$ (excluding H$_\epsilon$
to avoid the \ion{Ca}{ii}~H line), the \ion{He}{i} lines
$\lambda\lambda$ 4026~\AA, 4388~\AA, 4471~\AA, 4921~\AA\, and the
\ion{He}{ii} lines $\lambda\lambda$ 4542~\AA, 4686~\AA\ for the
helium-poor stars. For the helium-rich stars we also included the
\ion{He}{i} lines $\lambda\lambda$ 4713~\AA, 5015~\AA\ and 5044~\AA\
in the fit.  The results are given in Table~\ref{tab-par} for
NGC~2808 and Table \ref{tab-paro} for $\omega$ Cen.

\begin{figure}[h]
\vspace*{7cm}
\includegraphics{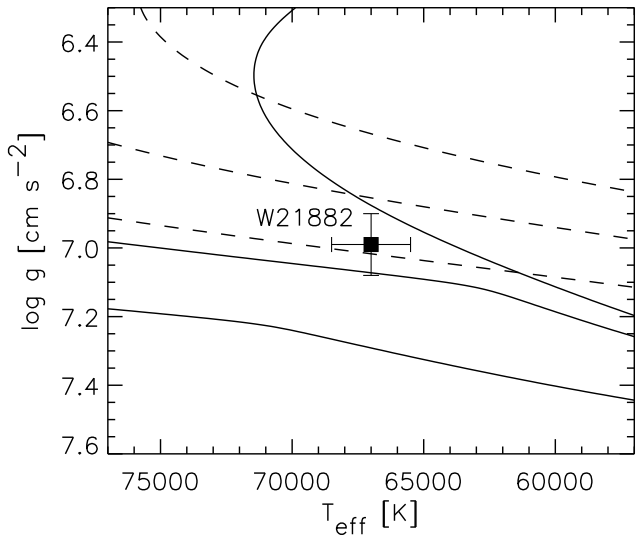}
\caption{Comparison of the stellar parameters for W21882 (solid square) 
with evolutionary tracks for helium (dashed curves) and
carbon-oxygen (solid curves) white dwarfs.  The masses of the tracks
are 0.423, 0.448 and 0.480~\Msolar\ for the helium
white dwarfs and 0.491, 0.494, and 0.529~\Msolar\ for
the carbon-oxygen white dwarfs in order of increasing gravity.
\label{W21882}}
\end{figure}

\begin{figure}[h]
\vspace*{8.5cm}
\includegraphics{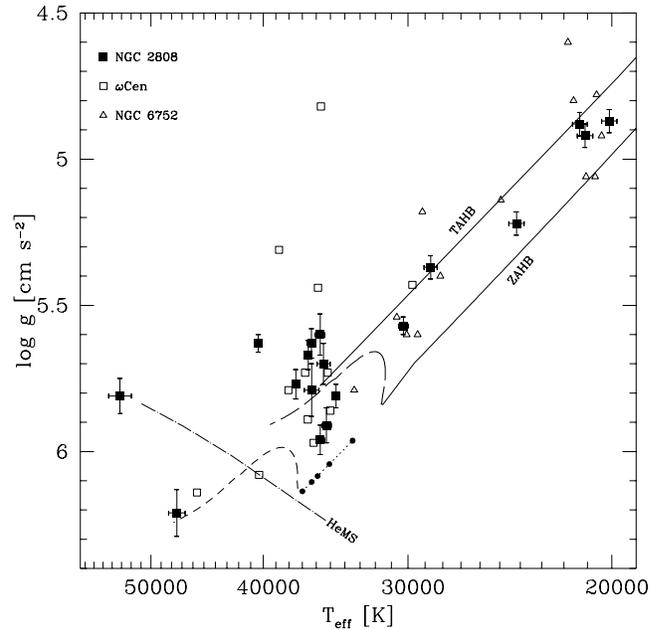}
\caption{Atmospheric parameters derived from the spectra of stars
along the blue tail/blue hook in NGC~2808 (filled squares,
Table~\ref{tab-par}) compared to HB evolutionary tracks. Also shown
are blue tail stars from NGC~6752 (open triangles, Moehler et
al. \cite{mosw00}) and the blue hook stars in $\omega$ Cen (open
squares, Table~\ref{tab-paro}).  The tracks for an early hot flasher
(long-dashed line) and a late hot flasher (short-dashed line) show the
evolution of such stars from the zero-age HB (ZAHB) towards helium
exhaustion in the core (terminal-age HB = TAHB). The solid lines mark
the canonical HB locus for [M/H] = $-$1.5 from Sweigart
(\cite{swei97}). The dotted line connects the series of ZAHB models
computed by adding a hydrogen-rich layer to the surface of the ZAHB
model of the late hot flasher.  The large dots mark -- with
decreasing temperature -- hydrogen layer masses of $0, 10^{-7},
10^{-6}, 10^{-5}, 10^{-4}$\Msolar.
\label{ngc2808_tg}}
\end{figure}

\begin{figure}[h]
\vspace*{8.5cm}
\includegraphics{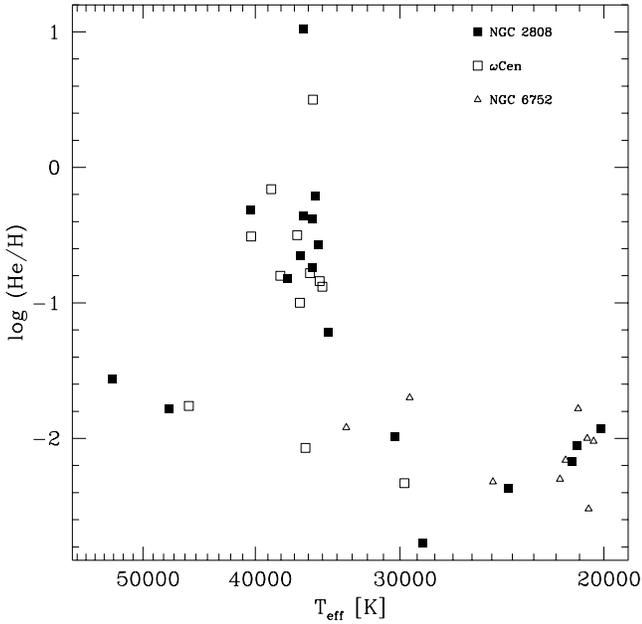}
\caption{Helium abundances vs. effective temperature for the stars
along the blue tail/blue hook in NGC~2808 (filled squares,
Table~\ref{tab-par}). Also shown are blue tail stars from NGC~6752
(open triangles, Moehler et al. \cite{mosw00}) and the blue hook stars
in $\omega$ Cen (open squares, Table~\ref{tab-paro}).
\label{ngc2808_the}}
\end{figure}

As can be seen from Table~\ref{tab-par} W21882 shows both a much
higher effective temperature (67,000~K) and surface gravity (\logg\ =
6.99) than all other stars.  These stellar parameters indicate that
W21882 may be a low-mass white dwarf, just in the mass range between
the most massive He white dwarfs and the least massive C/O white
dwarfs (cf. Fig~\ref{W21882}).  This suggests an absolute magnitude of
$M_V \approx$\magpt{6}{7}, which would place W21882 at a distance of
at most 4 kpc (zero reddening assumed), compared to 9.6 kpc for
NGC~2808. Unfortunately, the radial velocity estimates derived from
our spectra are too crude to determine cluster membership.  However,
as can be seen from Table~\ref{tab-par}, the model fit to W21882 has
by far the largest $\chi^2$ value of all our targets.  We examined the
three individual spectra to search for variability that might indicate
binarity or problems with the spectra, but did not find any such
evidence. Additional observations of this very hot star are needed to
better determine its cluster membership and evolutionary status and we
therefore omit it from all further discussion.  Figs.~\ref{ngc2808_tg}
and \ref{ngc2808_the} show the results for all other stars.

\begin{figure}[h]
\vspace*{8cm}
\includegraphics{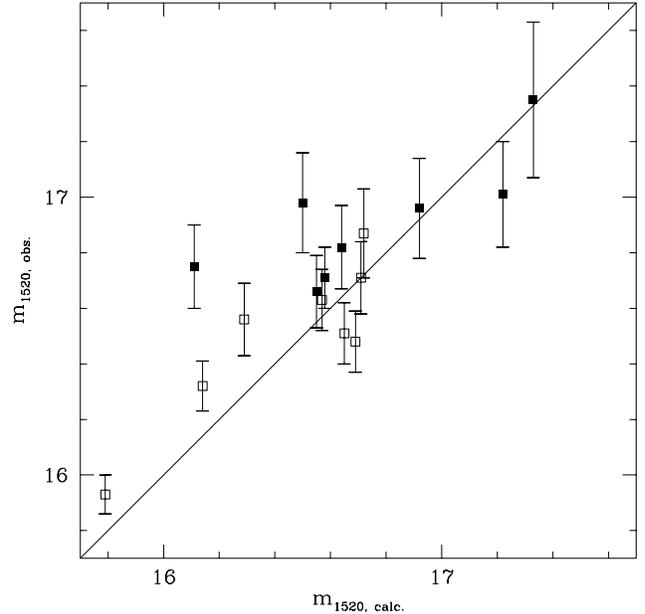}
\caption{Observed UIT magnitudes vs. UIT magnitudes calculated from
observed $V$ magnitudes and model spectra. The solid line marks
identity between observed and calculated UIT magnitudes. Filled squares mark
helium-rich stars, open squares mark helium-poor stars.\label{uit}}
\end{figure}

In order to verify if the atmospheric parameters can reproduce the UIT observations, we calculated UIT fluxes from the observed $V$
magnitudes, using the model spectra from the best fits. The model
spectra were first reddened by \ebv\ = \magpt{0}{18}. Then mean
fluxes through both the UIT 1520~\AA\ and the $V$ filters were
calculated to obtain the $m_{1520}-V$ colour, which was added to
the observed $V$ magnitude from Walker (\cite{walk99}) to obtain
the expected $m_{1520}$ magnitude.  The results are compared to the
UIT observations in Fig.~\ref{uit}.  The agreement is good for the
majority of the stars considering the low quality of the UIT
data.  The largest discrepancies are for the two helium-rich stars
(W6849 and W14040), which have much fainter observed UIT magnitudes than
predicted.  A possible cause for these discrepancies might be the
omission of carbon or other metal opacity in our model atmospheres,
which Lanz et al.\ (\cite{labr03}) have shown can affect both the
derived atmospheric parameters from optical spectra, and the
ultraviolet flux distribution. 


\subsection{Carbon abundances}
 Along with a large helium enhancement, the flash-mixing scenario
also predicts a strong surface carbon enhancement.  The flash-mixed
model computed by Cassisi et al.  (\cite{casc03}) is strongly enhanced in both
carbon (mass fraction 0.029) and nitrogen (mass fraction 0.007). The
corresponding solar mass fractions are 0.0028 and 0.00081 for C and N,
respectively. We see lines of \ion{C}{iii} $\lambda\lambda$ 4070,
4170, 4650~\AA\ in several of our spectra, but our low resolution and S/N
prevents a detailed abundance analysis.

To obtain a rough estimate of the carbon abundance, we began with the
H-He NLTE models used to derive Teff, log g and the helium abundance.
Keeping these parameters fixed we then computed five additional models
with CNO\footnote{Although oxygen should not be enhanced in
flash-mixed stars, we do not expect the enhanced oxygen in our models
to alter our estimate of the carbon abundance.} logarithmic mass
fractions relative to solar of $-1.5$, $-0.5$, $0.0$, $+0.5$, and
$+1.0$. As noted by Lanz et al. (\cite{labr03}), this approach is inconsistent
for large carbon abundances, in that the additional C opacity can
alter the atmospheric structure significantly, and therefore the
derived stellar parameters. The impact of this additional C
opacity has so far been explored only in the case of the field He-sdB
star PG1544$+$488. For this star Lanz et al. (\cite{labr03}) found that the
value of \logg\ given by NLTE models with a C abundance of 0.02 by
mass, as obtained by fitting the UV C III lines, was about 0.5 dex larger
than the value given by models without C.  Additional calculations are
needed to determine the carbon abundance at which it is possible to
neglect carbon in the model atmosphere and still derive accurate
stellar parameters.  We note, however, that we find a supersolar
carbon abundance in only one star (W20334), and so our crude approach
is likely adequate for the remaining stars.

The comparison between model spectra and observations resulted in a
carbon abundance compatible with the cluster abundance ([Fe/H] $=
-1.15$, Harris \cite{harr96}) except for the three most helium-rich
stars W14040, W18899, and W20334.  For W14040 (\loghe = $-0.31$) and
W18899 (\loghe = $-0.21$) we estimate logarithmic carbon mass fraction
relative to the Sun of $-$0.5 and 0.0, respectively. The most
helium-rich star, W20334 (\loghe = +1.02), shows \ion{C}{iii} lines at
$\lambda\lambda$ 4070~\AA\ and 4650~\AA, which are even stronger than
predicted by a model atmosphere with a CNO abundance of ten times
solar. In this case, the limitations of our two-step approach (see
above) become evident. As noted above,
a more accurate abundance determination for
this star requires an analysis with a model grid taking carbon opacity
into account for the determination of \teff\ and \logg\ (Lanz et
al. \cite{labr03}).

A similar estimate of the carbon abundances for the $\omega$ Cen stars
was not possible due to the poor S/N and resolution of these data.
Nevertheless, the strong \ion{C}{iii} features found in some of
the $\omega$ Cen spectra suggest an enhanced carbon abundance in some
of the blue hook stars.

\section{Discussion\label{sec-discuss}}

\begin{figure}[h]
\vspace*{8cm}
\includegraphics{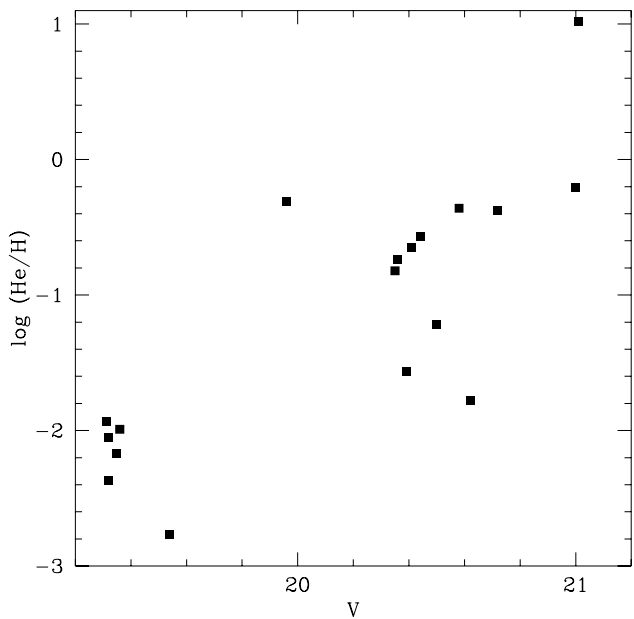}
\caption{Helium abundance vs. $V$ magnitudes. The gap in the
colour-magnitude diagram is located at $V\approx$ \magn{20}, and
\loghe\ = $-1$ corresponds to solar helium abundance.\label{gap}}
\end{figure}

Our analysis of blue tail and blue hook stars in NGC~2808 shows that
these stars form two different groups according to their helium
abundance (cf. Fig.~\ref{gap}): All stars brighter than the gap at
V$\approx$\magn{20} are helium-poor, although they cover a large
temperature range (cf. Fig.~\ref{ngc2808_the}). The stars fainter
than the gap at V$\approx$\magn{20}, on the other hand, show mostly
super-solar abundances, with three exceptions: W8750, W9863, and
W23198. Fig.~\ref{ngc2808_the} shows that the helium-rich stars
cluster between 35,000~K and 40,000~K, well beyond the hot end of the
canonical EHB. We will now discuss the evolutionary status of the
helium-poor and helium-rich stars in more detail.

\subsection{Helium-Poor Stars}

As can be seen from Fig.~\ref{ngc2808_tg}, the six helium-poor stars
cooler than 31,000~K occupy a range in effective temperature and
surface gravity similar to the blue tail stars in NGC~6752. Their
helium abundances are also similar to those observed in NGC~6752,
indicating that the gravitational settling of helium has been equally
efficient in both clusters.  Thus, the helium-poor stars brighter than
the gap at V$\approx$\magn{20} agree with the predictions of canonical
HB evolution.

\begin{figure}[h]
\vspace*{7cm} 
\includegraphics{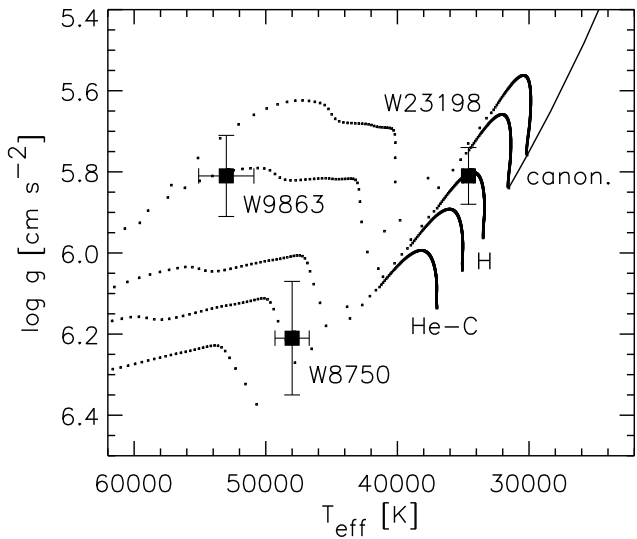}
\caption{Comparison of the stellar parameters for W8750, W9863, and W23198
 (solid squares) with EHB and post-EHB evolutionary tracks.  Each
track is represented by a series of points separated by a time
interval of 5$\cdot$10$^5$ yr in order to indicate the rate of
evolution.  The solid line in the upper righthand corner denotes the
canonical ZAHB.  The two lowest gravity tracks are for canonical,
i.e., unmixed models with hydrogen-rich envelopes.  The two
intermediate tracks  labeled H) are for flash-mixed models with
a hydrogen-rich surface layer containing either 10$^{-5}$ \Msolar\
(higher gravity track) or 10$^{-4}$~\Msolar (lower gravity track). The
highest gravity track  labeled He-C) is for a flash-mixed model
with no envelope hydrogen.
\label{postHB}}
\end{figure}

The two helium-poor stars hotter than 40,000~K, W8750 and W9863,
appear to be in a post-HB evolutionary state This can be seen in
Fig.~\ref{postHB}, where we compare the stellar parameters of these
stars with HB and post-HB evolutionary tracks for canonical models,
flash-mixed models with a thin hydrogen surface layer, and flash-mixed
models with no envelope hydrogen. W9863 appears to lie along the
hottest canonical post-HB track at a point where the evolution is
relatively slow.  However, things are less simple for W8750.  While it
lies close to the post-HB track of a late hot flasher with a
10$^{-5}$ \Msolar\ hydrogen surface layer, it also lies at a point
where the evolution is rapid.  Moreover, it is difficult to understand
why W8750 has a helium-poor surface when the other stars with possible
flash mixing are much more helium-rich. In order to place it on the
least massive canonical track (where evolution is slow in this
temperature range), its gravity would have to be overestimated
by about 0.4 dex, which is about 3$\sigma$.

W23198 is a less clear case: In Figs.~\ref{ngc2808_tg} and
\ref{ngc2808_the} it lies at the cool end of the population of
helium-enriched stars, but well separated from the canonical EHB
stars. Judging from its position in Fig. 9, W23198 may be a late hot
flasher with very high hydrogen content and/or effective diffusion or
an example for shallow mixing (see below).

\subsection{Helium-Rich Stars}

The parameters of the helium-rich stars provide support to the late
hot flasher scenario: Figs.~\ref{ngc2808_tg} and
\ref{ngc2808_the} show that a gap between
$\approx$31,000~K to $\approx$35,000~K separates the helium-rich from the 
helium-poor stars, as predicted by the
flash-mixing scenario. 
The HB track for the early hot flasher in Fig.~\ref{ngc2808_tg} passes
through the temperature gap,  but the evolution is then very fast,
making a contamination of the temperature gap by such stars 
unlikely (Moehler et al. \cite{mosw02}).
In addition, Fig.~\ref{ngc2808_the} shows the expected
clustering of the helium-rich stars between 35,000~K and 40,000~K.

Contrary to the predictions of Brown et al. (\cite{brsw01}) and
Cassisi et al. (\cite{casc03}), but consistent with our previous
results for $\omega$ Cen, the atmospheres of the blue hook stars still
show some hydrogen. This result has been discussed by Cassisi et
al. (\cite{casc03}), who find that to reproduce even the highest
observed helium abundances, they have to reduce the efficiency
of the flash mixing by a factor of about 20,000. However, this
reduction only applies if the observed helium abundances reflect the
actual helium abundances in the envelopes of the blue hook stars.

 This apparent discrepancy could be explained if some residual
hydrogen survived flash mixing and later diffused outward, thus
producing a thin hydrogen-rich layer at the surfaces of the blue hook
stars. Such diffusive processes are believed to be responsible for the
low helium abundances of the sdB stars and are estimated to operate on
a time scale much shorter than the HB lifetime.  It is unclear,
however, if diffusion will turn a helium-rich star into a helium-poor
one. Groth et al. (\cite{grku85}) found that atmospheric convection
(which would work against diffusion) can exist in hot subdwarfs if the
helium abundance is sufficiently high. Also mass loss might
affect the atmospheric abundances of the blue hook stars by reducing
the efficiency of the hydrogen diffusion. The low helium abundances
observed for canonical EHB stars, however, suggest that mass loss does
not prevent the formation of a hydrogen surface layer. The range in
the hydrogen abundances of the blue hook stars might indicate that
varying amounts of hydrogen survive flash mixing or that the
efficiency of diffusion differs from star to star. In any case the
high helium abundances observed in some of the blue hook stars would
be difficult to understand if their atmospheres were not enriched in
helium during the helium core flash. The increase in the mean
atmospheric helium abundance with increasing effective temperature is
also consistent with flash mixing.  As can be seen in
Fig.~\ref{ngc2808_tg}, adding a hydrogen layer to the surface of a
late hot flasher moves the track towards lower temperatures and
gravities (see Moehler et al. \cite{mosw02} for more details).  The
addition of a hydrogen layer of $<$ 10$^{-4}$~M$_\odot$ would actually
improve the agreement between the predicted and observed temperatures
of the blue hook stars while at the same time preserving the
temperature gap between these stars and the canonical EHB stars.

Another possibility for explaining the hydrogen abundances of the blue
hook stars has been discussed by Lanz et al. (\cite{labr03}), who
found that there are two types of flash mixing: ``deep" and
``shallow", depending on how far the envelope hydrogen is mixed into
the core during a late helium-core flash.  During deep mixing the
envelope hydrogen is mixed all the way into the high temperature
regions near the site of the helium flash and therefore rapidly
burned.  This is the type of mixing discussed by Brown et
al. (\cite{brsw01}) and Cassisi et al. (\cite{casc03}).  A blue hook
star following deep mixing will be helium- and carbon- rich with very
little hydrogen left in its atmosphere.  In contrast, during shallow
mixing the envelope hydrogen is only mixed with the outer layers of
the core.  Since the temperatures within these layers are too low for
proton-capture nucleosynthesis, all of the envelope hydrogen will
survive the flash-mixing phase.  The surface composition of a blue
hook star following shallow mixing will therefore remain
hydrogen-rich, although its surface composition will be diluted by
helium- and carbon-rich material from the core.  Unfortunately the
calculations of Lanz et al. (\cite{labr03}) indicate that shallow
mixing only occurs over a very narrow range of mass loss in metal-poor
stars.  Thus, while shallow mixing may be a viable explanation for the
residual hydrogen found in some field He-rich sdB stars (Lanz et
al. \cite{labr03}), it is probably of minor importance for the blue
hook stars in NGC 2808.  A detailed study of metal abundances might be
able to distinguish the signatures of shallow mixing from those of
diffusion.

\section{Conclusions}

The high temperatures and high helium abundances reported here for the
blue hook stars in NGC~2808 and $\omega$ Cen provide general support
for the flash-mixing scenario of Brown et al. (\cite{brsw01}). The three most helium-rich stars in NGC~2808 also show evidence
for carbon enrichment.  However, all of our targets show some hydrogen
in their atmospheres, and only one target in each cluster has \loghe
$> 0$.  This is partially a selection effect, as our targets are
preferentially the brighter blue hook stars, and the most helium-rich
star in each cluster is either the faintest (W20334 in NGC~2808) or
second-faintest (D10763 in $\omega$ Cen).  However, it is also likely
that some hydrogen survives a late hot flash, and subsequently
diffuses to the surface during the HB phase.

Moehler et al. (\cite{mohe97a}) reported an isolated helium-rich sdB
star in M15 with the expected properties (\teff\ = 36,000~K, M$_V$ =
\magpt{+4}{7}, \loghe = +0.82) of a flash-mixed star.  However,
significant numbers of blue hook stars have been found only in the
most massive Galactic globular clusters: $\omega$ Cen (M$_V$ =
\magpt{-10}{29}), NGC~2808 (M$_V$ = \magpt{-9}{39}), NGC~6388 (M$_V$ =
\magpt{-9}{42}), and possibly NGC~6441 (M$_V$ = \magpt{-9}{64}), all
data for M$_V$ from Harris (\cite{harr96}). Other globular cluster in
the same mass regime are 47~Tuc (M$_V$ = \magpt{-9}{42}), NGC~2419
(M$_V$ = \magpt{-9}{58}), and NGC~6715 (M$_V$ = \magpt{-10}{01}). The
optical colour-magnitude diagrams of NGC~2419 (Harris et
al. \cite{habe97}) and NGC~6715 (Momany, priv. comm., Rosenberg et
al. \cite{rore03}) show stars faint enough to be blue hook stars. The
situation for 47~Tuc is unclear due to contamination by the Small
Magellanic Cloud.

 Apparently the number of blue hook stars depends on the total mass of
the globular cluster and not on the number of other hot HB stars (see
Rosenberg et al. \cite{rore03} for an independent discussion of this
effect). For instance, NGC~6752, which is a factor of 100 less
luminous than NGC~2808 and has a very long and well populated blue
tail, contains 59 EHB stars with \magpt{3}{0} $\le M_V \le$
\magpt{4}{5} (photometry of Buonanno et al. \cite{buca86}), compared
to 34 such stars observed in NGC~2808 (photometry from Walker
\cite{walk99}).  Even more EHB stars can be seen in the data of Momany
et al. (\cite{mopi02}) for NGC~6752.  However, while NGC~2808 has 28
blue hook stars with \magpt{4}{5} $\le M_V \le$ \magpt{5}{5}, {\em no}
such stars are found in NGC~6752.  Likewise, the UV colour-magnitude
diagram of Brown et al. (\cite{brsw01}) shows that the population of
blue hook stars dominates over the population of EHB stars in NGC~2808
(46 vs. 29 stars, respectively).  As the mass loss ranges populating
the hot end of the canonical extreme HB and the blue hook region are
not too different, it is hard to understand why globular clusters with
large numbers of canonical EHB stars do not show at least some blue
hook stars as well.

\begin{acknowledgements} 
We appreciate the efforts of the staff at Paranal in performing the
observations. We are grateful to Uli Heber for his remarks about
convection and to Ralf Napiwotzki for his model atmospheres, fit routines
and help with W21882. We thank the referee S. Cassisi for
valuable suggestions and a very timely referee report.
SM thanks the DFG for a travel
grant (444 USA 111/4/03). 
\end{acknowledgements}

\end{document}